\documentclass[aps,pra,reprint,showpacs,superscriptaddress]{revtex4-1}
\usepackage{amsmath,amssymb,mathrsfs,amsfonts}
\usepackage{epsfig}
\usepackage{bm}
\usepackage{color}
\usepackage[dvipdfm,
            pdfstartview=FitH,
            CJKbookmarks=true,
            bookmarksnumbered=true,
            bookmarksopen=true,
            linktocpage=true,
            colorlinks=true, 
            pdfborder=001,   
            citecolor=blue,  
            urlcolor=blue,   
            linkcolor=blue,  
            anchorcolor=blue,
            ]{hyperref}      

\begin{document}
  \title{Geometry of quantum evolution in a nonequilibrium environment}
  \author{Xiangji Cai}
   \email{xiangjicai@foxmail.com}
  \affiliation{School of Science, Shandong Jianzhu University, Jinan 250101, China}
  \author{Ruixuan Meng}
  \affiliation{School of Science, Shandong Jianzhu University, Jinan 250101, China}
  \author{Yanhui Zhang}
  \email{yhzhang@sdnu.edu.cn}
  \affiliation{School of Physics and Electronics, Shandong Normal University, Jinan 250014, China}
  \author{Lifei Wang}
  \affiliation{School of Science, Shandong Jiaotong University, Jinan 250357, China}

\begin{abstract}
  We theoretically study the geometric effect of quantum dynamical evolution in the presence of a nonequilibrium noisy environment.
  We derive the expression of the time dependent geometric phase in terms of the dynamical evolution and the overlap between the time evolved state and initial state.
  It is shown that the frequency shift induced by the environmental nonequilibrium feature plays a crucial role in the geometric phase and evolution path of the quantum dynamics.
  The nonequilibrium feature of the environment makes the length of evolution path becomes longer and reduces the dynamical decoherence and non-Markovian behavior in the quantum dynamics.
\end{abstract}

\maketitle

\section{Introduction}
\label{sec:intr}
The global phase related to the
dynamical evolution of a quantum system contains a
gauge-invariant component, namely, the geometric phase
which depends only on the geometry of the path traversed
by the system during the quantum evolution~\cite{ProcIndAcadSciA44.247,ProcRSocA392.45,Shaperebook,RevModPhys.64.51,Bohmbook}.
Due to the fact that a quantum system unavoidably interacts
with its environment and undergoes decoherence, much
extensive attention has been paid to theoretical investigations
on the geometric phase in open quantum systems
under nonunitary dynamics~\cite{PhysRevLett.58.1593,PhysRevLett.60.2339,PhysRevLett.90.160402,PhysRevLett.93.080405,PhysRevA71.032106,PhysRevLett.95.157203,PhysRevLett.96.150403,
PhysRevA71.044101,PhysRevA73.052103,PhysRevA73.062104,PhysRevA74.042311,PhysRevA75.032103,PhysRevLett.105.240406,
PhysRevA83.052121,PhysRevA85.032105,PhysRevA86.064104,PhysRevLett.108.170401,PhysRevA89.012110,EurophysLett.118.50003,PhysRevA98.052117,SciRep.8.9852}.
The geometric phase
associated with quantum evolution has been observed
and measured in a variety of experiments~\cite{PhysRevLett.57.933,PhysRevLett.57.937,PhysRevLett.59.251,PhysRevLett.60.1218,Science318.1889,
PhysRevLett.91.100403,PhysRevA76.042121,PhysRevLett.100.177201,PhysRevA80.054101,PhysRevA95.042345}, and it
has demonstrated that its geometric feature has potential
applications in studying quantum phase transition and
realizing geometric quantum computation~\cite{PhysRevLett.95.157203,PhysRevLett.96.077206,Nature403.869,PhysRevB67.121307,PhysRevLett.90.028301}.
The investigation on the geometry in the dynamical evolution
of an open quantum system is crucial for further understanding
the origins of decoherence, quantum-classical transition and so on.

With the development of experimental techniques to
control and manipulate quantum systems at different time
scales and energy ranges, the study of non-Markovian behavior
in the dynamical evolution of open quantum systems
has increasingly evolved into an attractive research field~\cite{Breuerbook,PhysRevLett.103.210401,PhysRevLett.105.050403,PhysRevLett.109.170402,PhysRevLett.112.120404,
RepProgPhys.77.094001,RevModPhys.88.021002,RevModPhys.89.015001,PhysRevLett.118.140403,JStatMech.2016.054016,PhysRevA97.042126,EurophysLett.118.20005}.
Meanwhile, the non-Markovian effect of
the dynamics in open quantum systems on the geometric
phase has been well studied in equilibrium environments
with both Markov and stationary statistical properties~\cite{EurophysLett.82.50001,PhysRevA81.022120,PhysRevA82.052111,PhysRevA90.062133}.
As a matter of fact, there are many significant
situations where a nonequilibrium environment has an essential
influence on the dynamical evolution of a quantum
system. For example, in transient and ultrafast processes
in physical or biological systems, some dynamical behavior
occurs on sufficiently short time scales, and there may
be no chance for the environmental initial nonequilibrium
states induced by the coupling between the system and
environment to reach equilibrium rapidly~\cite{JChemPhys.133.241101,JPhysB45.154008,PhysRevLett.112.246401,PhysRevLett.115.257001}.

The environment with nonstationary statistics has been
taken into extensive consideration, corresponding physically
to impulsively environmental excited phonons out of
thermal equilibrium states initially. It has drawn much attention
in the study of dynamical decoherence, geometric
phases and quantum speed limits of open quantum systems
in nonequilibrium environments and quantum measurements
detected by a voltage-biased quantum point
contact (QPC) or a single-electron transistor~\cite{PhysRevA78.032105,JChemPhys.139.024109,PhysRevA87.032338,PhysRevA91.042111,PhysRevA94.042110,
PhysRevA95.023610,PhysRevA95.052104,JChemPhys.149.094107,PhysRevB96.235417,SciRep.5.11726}.
Given that a quantum system may interact with a composite
or structured environment, where the coupling between
the sub-environments plays an essential role in the dynamical
evolution of the quantum system, not only the system
dynamics but the statistical properties of the environment display the non-Markovian feature, namely, the memory
effect of the environmental noise~\cite{PhysRep.88.207,PhysRevE50.2668,EurophysLett.118.60002}.
It has been
shown that the nonequilibrium feature of the environment
gives rise to a Lamb-type renormalization of the intrinsic
energy levels which contributes additionally to the unitary
dynamical evolution of the quantum system and that
the environment non-Markovian feature may not result in
non-Markovian behavior in the dynamics of the quantum
system~\cite{PhysRevA94.042110,JChemPhys.149.094107}.

In this paper, we theoretically study the geometric effect
of evolution of a two-level quantum system coupled
to a nonequilibrium noisy environment. Based on the
quantum master equation and geometric phase defined
for nonunitary evolution, we derive the time-dependent
geometric phase composed of the contributions from the
dynamical evolution and the overlap between the time
evolved state and initial state. We discuss how the environmental
nonequilibrium feature influences the geometry
of quantum evolution and explore the mechanism for
the geometric effect of quantum dynamical evolution in
a nonequilibrium noisy environment. It is shown that
the renormalization of the intrinsic energy of the system,
namely, the frequency shift induced by the nonequilibrium
feature of the environment has a significant impact on
both the geometric phase and the evolution path of the
dynamics.

\section{Theoretical framework}
\label{sec:theo}
We consider a two-level
quantum system interacting with a nonequilibrium noisy
environment. The environmental effect leads to the intrinsic
energy of the quantum system driven linearly by
a nonstationary and non-Markovian stochastic noise process.
In an equilibrium environment, when the interaction
Hamitonian is commutative with the intrinsic Hamiltonian
of the system, the pure decoherence process does not cause
the energy renormalization, whereas in a nonequilibrium
environment, it gives rise to the renormalization of the intrinsic
energy levels due to the nonstationary statistical
properties of the environmental noise~\cite{PhysRevA94.042110,JChemPhys.149.094107}.

For the pure decoherence in a nonequilibrium environment,
the uncontrolled environmental degrees of freedom
give rise to the stochastic fluctuations in the Hamiltonian
of the quantum system as~\cite{JPhysSocJpn.9.316,JPhysSocJpn.9.935}
\begin{equation}
\label{eq:redevol}
H(t)=H_{0}+H_{\xi}(t)=\frac{\hbar}{2}\omega_{0}\sigma_{z}+\frac{\hbar}{2}\xi(t)\sigma_{z},
\end{equation}
where $H_{0}$ is the intrinsic Hamiltonian of the quantum system, $H_{\xi}(t)$ is the stochastic fluctuating term caused by the interaction with the environment, $\sigma_z$ is the Pauli matrix, $\omega_{0}$ denotes the intrinsic frequency difference between the excited state $|e\rangle$ and ground state $|g\rangle$, and $\xi(t)$ represents the environmental noise exhibiting both nonstationary and non-Markovian features.
The dynamical evolution for the total density matrix yields the Liouville equation
\begin{equation}
\label{equS2}
  \frac{\partial}{\partial t}\rho\bm(t;\xi(t)\bm)=-\frac{i}{\hbar}[H(t),\rho\bm(t;\xi(t)\bm)].
\end{equation}
The reduced density matrix of the quantum system can be derived by taking an ensemble average over the environmental noise as $\rho(t)=\left\langle\rho\bm(t;\xi(t)\bm)\right\rangle$.
In a nonequilibrium environment, the dynamical evolution for the reduced density matrix of the quantum system is governed by a time-local master equation as~\cite{PhysRevA95.052104,JChemPhys.149.094107}
\begin{equation}
\label{eq:redevol}
  \frac{d}{dt}\rho(t)=-\frac{i}{2}[\omega_{0}-s(t)][\sigma_{z},\rho(t)]
                                   +\frac{1}{2}\gamma(t)[\sigma_{z}\rho(t)\sigma_{z}-\rho(t)].
\end{equation}
Due to the nonstationary statistical properties of the environmental noise, the decoherence factor $F(t)=|F(t)|e^{i\phi(t)}$ is a complex time dependent function with the modulus $|F(t)|$ and the argument $\phi(t)$ and it is employed to quantify the coherence evolution of the system initially prepared in the superposition state in the basis $\{|e\rangle,|g\rangle\}$.
The quantum evolution of the reduced density matrix is closely associated with the time dependent frequency shift $s(t)$ and decoherence rate $\gamma(t)$ which are defined, respectively, as
\begin{equation}
\label{eq:frshdera}
\begin{split}
  s(t)=&-\mathrm{Im}\left[\frac{dF(t)/dt}{F(t)}\right]=-\frac{d}{dt}\phi(t),\\
  \gamma(t)=&-\mathrm{Re}\left[\frac{dF(t)/dt}{F(t)}\right]=-\frac{1}{|F(t)|}\frac{d}{dt}|F(t)|,
\end{split}
\end{equation}
where the frequency shift $s(t)$ is used to identify the decoherence processes in equilibrium and nonequilibrium environments and the decoherence rate $\gamma(t)$ is related closely to the exchange of information between the system and environment, namely, non-Markovian behavior in the system dynamics.

To quantify the non-Markovian effect in the dynamical evolution, numerous measures for non-Markovianity in open quantum systems have been proposed based on different quantification of the distinguishability between quantum states, different divisibility properties of the quantum dynamical map
and some other quantities related to the concepts in quantum information
which exhibit a monotonic or an oscillating behavior in time~\cite{PhysRevLett.103.210401,PhysRevLett.105.050403,PhysRevLett.109.170402,PhysRevLett.112.120404}.
A most widespread measure for quantification of non-Markovianity in the dynamics of open quantum systems is based on the general notion of distinguishability of quantum states:
when the distinguishability decreases, the information flows from the system into the environment,
while an increase of the trace distance signifies a flow of information from the environment
back into the system.
Based on this measure, the non-Markovianity in the dynamical evolution of the quantum system can be generally defined by~\cite{PhysRevLett.103.210401}
\begin{equation}
\label{eq:nonMar}
 \mathcal{N}=\max\limits_{\rho_{1,2}(0)}\int_{\sigma>0}\sigma\textbf{(}t,\rho_{1,2}(0)\textbf{)}dt.
\end{equation}
where $\sigma\textbf{(}t,\rho_{1,2}(0)\textbf{)}=dD\textbf{(}\rho_{1}(t),\rho_{2}(t)\textbf{)}/dt$
denotes the rate of change of the trace distance
\begin{equation}
\label{eq:tradis}
 D(\rho_{1},\rho_{2})=\frac{1}{2}\mathrm{tr}|\rho_{1}-\rho_{2}|,
\end{equation}
with $|A|=\sqrt{A^{\dag}A}$ being the modulus of an operator $A$ and the bound $0\leq D\leq1$.
It is worth pointing out that it is more convenient to employ an insensitive measure constructed in Ref.~\cite{EurophysLett.118.20005} to quantify the degree of non-Markovian behavior in the dynamics
arising from the possible overestimate of fluctuations in the trace distance, such as in the case driven by an external field.
The maximum of the trace distance difference in Eq.~\eqref{eq:nonMar} can be obtained by taking optimization over all pairs of initial states, and thus the time-dependent non-Markovianity can be written as~\cite{RevModPhys.88.021002}
\begin{equation}
\label{eq:non-Mar}
 \mathcal{N}(t)=-\int_{0\ \gamma(\tau)<0}^{t}\gamma(\tau)|F(\tau)|d\tau,
\end{equation}
where the optimal pairs of initial states are chosen as $\rho_{1,2}(0)=\frac{1}{2}(|e\rangle\pm|g\rangle)(\langle e|\pm\langle g|)$.

We consider the case that the environmental noise is subject to a nonstationary non-Markovian random telegraph process.
The amplitude of the noise process jumps randomly with the switching rate $\lambda$ between the values $\pm\nu$, and its nonstationary and non-Markovian features are characterized by the nonequilibrium parameter $a$ and an exponential form of memory kernel $K(t-t')=\kappa e^{-\kappa(t-t')}$ with the decay rate $\kappa$, respectively. The environment is in equilibrium for the nonequilibrium parameter $a=0$ when the environmental noise exhibits the stationary feature, and the environment is memoryless when the environmental noise is Markovian for the decay rate $\kappa\rightarrow\infty$~\cite{PhysRevA94.042110,JChemPhys.149.094107}.
In the presence of nonstationary non-Markovian random telegraph noise, the decoherence factor for
the quantum system can be exactly written in the analytical expression~\cite{PhysRevA94.042110}
\begin{equation}
\label{eq:defac}
\begin{split}
   F(t)=&\mathscr{L}^{-1}[\mathcal{F}(p)],\\
   \mathcal{F}(p)=&\frac{p^{2}+\kappa p+2\kappa\lambda+ia\nu(p+\kappa)}{p^{3}+\kappa p^{2}+(2\kappa\lambda+\nu^{2})p+\kappa\nu^{2}},
\end{split}
\end{equation}
where $\mathscr{L}^{-1}$ denotes the inverse Laplace transform, and the initial conditions of the decoherence factor are given by $F(0)=1$ and $\phi(0)=0$.
The nonequilibrium feature of the environment only influences the imaginary component of the decoherence factor on account of its effect on the renormalization of the intrinsic energy levels of the quantum system.
It is worth mentioning that the decoherence factor $F(t)$ is a real time dependent function and there is no frequency shift $s(t)=0$ for the equilibrium case.

The time dependent geometric phase for the quantum system under nonunitary dynamical evolution has been derived as~\cite{PhysRevLett.93.080405}
\begin{equation}
\label{eq:geopha1}
\begin{split}
   \Phi_{g}(t)=&\arg\Bigg\{\sum_{k}\sqrt{\epsilon_{k}(0)\epsilon_{k}(t)}\langle\Psi_{k}(0)|\Psi_{k}(t)\rangle\\
               &\times\exp\bigg[-\int_{0}^{t}\Big\langle\Psi_{k}(\tau)\Big|\frac{\partial}{\partial\tau}\Big|\Psi_{k}(\tau)\Big\rangle d\tau\bigg]\Bigg\},
\end{split}
\end{equation}
where $\epsilon_{k}(t)$ and $|\Psi_{k}(t)\rangle$ are the $k$th time-dependent eigenvalues and eigenvectors of the reduced density matrix $\rho(t)$, respectively.
Due to the environmental effect, the evolution of the quantum system is no longer cyclic and the system evolves along a quasicyclic path depending on the evolution time.

We express the state of the quantum system in terms of the Bloch vector as
\begin{equation}
\label{eq:Blovec}
 \rho(t)=\frac{1}{2}\left[I_{2}+\vec{r}(t)\cdot\vec{\sigma}\right],\ |\vec{r}(t)|\leq1,
\end{equation}
where $I_{2}$ is the $2\times2$ identity matrix, $\vec{\sigma}=(\sigma_{x},\sigma_{y},\sigma_{z})$ is the vector of Pauli matrices and $\vec{r}(t)=(r_{x}(t),r_{y}(t),r_{z}(t))$ denotes a real vector with
\begin{equation}
\label{eq:Blocom}
\begin{split}
  r_{x}(t)=&\mathrm{tr}[\sigma_{x}\rho(t)]=\rho_{eg}(t)+\rho_{ge}(t),\\
  r_{y}(t)=&\mathrm{tr}[\sigma_{y}\rho(t)]=i[\rho_{eg}(t)-\rho_{ge}(t)],\\
  r_{z}(t)=&\mathrm{tr}[\sigma_{z}\rho(t)]=\rho_{ee}(t)-\rho_{gg}(t).
\end{split}
\end{equation}
The state $\rho(t)$ is pure if and only if $|\vec{r}(t)|=1$, otherwise, mixed.
Based on Eq.~\eqref{eq:redevol}, the components of the Bloch vector $\vec{r}(t)$ satisfy the evolution
\begin{equation}
\label{eq:Bloevol}
\begin{split}
  \frac{d}{dt}r_{x}(t)&=-\gamma(t)r_{x}(t)-[\omega_{0}-s(t)]r_{y}(t),\\
  \frac{d}{dt}r_{y}(t)&=[\omega_{0}-s(t)]r_{x}(t)-\gamma(t)r_{y}(t),\\
  \frac{d}{dt}r_{z}(t)&=0.
\end{split}
\end{equation}

Consequently, the time-dependent eigenvalues of the reduced density matrix in terms of the Bloch vector can be expressed as
\begin{equation}
\label{eq:eigval1}
 \epsilon_{\pm}(t)=\frac{1}{2}[1\pm|\vec{r}(t)|],
\end{equation}
and the corresponding instantaneous eigenvectors can be written as
\begin{equation}
\label{eq:eigvec}
 |\Psi_{\pm}(t)\rangle=C_{\pm e}(t)|e\rangle+C_{\pm g}(t)|g\rangle.
\end{equation}
with the expressions of the time dependent complex coefficients
\begin{equation}
\label{eq:comcoe}
\begin{split}
 C_{\pm e}(t)=&\frac{r_{x}(t)-ir_{y}(t)}{\sqrt{[2\epsilon_{\pm}(t)-1-r_{z}(t)]^{2}+r_{x}^{2}(t)+r_{y}^{2}(t)}},\\
 C_{\pm g}(t)=&\frac{2\epsilon_{\pm}(t)-1-r_{z}(t)}{\sqrt{[2\epsilon_{\pm}(t)-1-r_{z}(t)]^{2}+r_{x}^{2}(t)+r_{y}^{2}(t)}}.
\end{split}
\end{equation}
If the initial state is pure, e.g., $|\vec{r}(0)|=1$, the smaller eigenvalue $\epsilon_{-}(t)$ makes no contribution to the geometric phase defined in Eq.~\eqref{eq:geopha1} since $\epsilon_{-}(0)=0$ at time $t=0$.
For the general case that the system is prepared in an arbitrary initial state, we can, based on Eq.~\eqref{eq:geopha1}, rewrite the geometric phase as
\begin{equation}
\label{eq:geopha2}
 \Phi_{g}(t)=\arg\big[r_{+}(t)e^{i\varphi_{+}(t)}e^{i\psi_{+}(t)}+r_{-}(t)e^{i\varphi_{-}(t)}e^{i\psi_{-}(t)}\big],
\end{equation}
where we have used the definitions
\begin{equation}
\label{eq:coe}
\begin{split}
 r_{\pm}(t)=&\left|\sqrt{\epsilon_{\pm}(0)\epsilon_{\pm}(t)}\langle\Psi_{\pm}(0)|\Psi_{\pm}(t)\rangle\right|,\\
 \varphi_{\pm}(t)=&\arg\langle\Psi_{\pm}(0)|\Psi_{\pm}(t)\rangle,\\
 \psi_{\pm}(t)=&i\int_{0}^{t}\Big\langle\Psi_{\pm}(\tau)\Big|\frac{\partial}{\partial\tau}\Big|\Psi_{\pm}(\tau)\Big\rangle d\tau\\
              =&\int_{0}^{t}\Big[\omega_{0}+\frac{d}{d\tau}\phi(\tau)\Big]|C_{\pm e}(\tau)|^{2}d\tau.
\end{split}
\end{equation}
It is obvious that the geometric phase for the system evolving from an arbitrary initial state is defined as a sum over the phase factors with the weights related closely to the time-dependent eigenvalues and eigenvectors of the reduced density matrix.

For simplicity, we assume that the system is initially prepared in the pure eigenstate
\begin{equation}
\label{eq:inista}
 |\Psi(0)\rangle=\cos\frac{\theta}{2}|e\rangle+\sin\frac{\theta}{2}|g\rangle,
\end{equation}
with the initial values of the components of the Bloch vector $r_{x}(0)=\sin\theta$, $r_{y}(0)=0$ and $r_{z}(0)=\cos\theta$.
After time $t$, the state of the quantum system evolves to
\begin{equation}
\label{eq:finsta}
 |\Psi(t)\rangle=e^{-i\omega_{0}t-i\phi(t)}\cos\theta_{+}(t)|e\rangle+\sin\theta_{+}(t)|g\rangle,
\end{equation}
where the time dependent real coefficients satisfy
\begin{equation}
\label{eq:reacoe}
\begin{split}
 \cos\theta_{+}(t)=&\frac{\sin\theta|F(t)|}{\sqrt{\big[2\epsilon_{+}(t)-1-\cos\theta\big]^{2}+\sin^{2}\theta|F(t)|^{2}}},\\
 \sin\theta_{+}(t)=&\frac{2\epsilon_{+}(t)-1-\cos\theta}{\sqrt{\big[2\epsilon_{+}(t)-1-\cos\theta\big]^{2}+\sin^{2}\theta|F(t)|^{2}}},
\end{split}
\end{equation}
with the larger eigenvalue $\epsilon_{+}(t)$ of the reduced density matrix which only makes a contribution to the geometric phase
\begin{equation}
\label{eq:eigval2}
 \epsilon_{+}(t)=\frac{1}{2}\left[1+\sqrt{\cos^{2}\theta+\sin^{2}\theta|F(t)|^{2}}\right].
\end{equation}
Thus, the time dependent geometric phase in Eq.~\eqref{eq:geopha2} can be expressed as
\begin{equation}
\label{eq:geopha3}
   \Phi_{g}(t)=\Phi_{P}(t)+\Phi_{e}(t),
\end{equation}
which contains the contribution arising from the overlap between the time evolved state $|\Psi(t)\rangle$ and initial state $|\Psi(0)\rangle$, namely, the Pancharatnam relative phase~\cite{ProcIndAcadSciA44.247,PhysRevLett.93.080405}, which can be written by
\begin{equation}
\label{eq:phaove}
\begin{split}
   \Phi_{P}(t)&=\arg\langle\Psi(0)|\Psi(t)\rangle\\
              &=-\arctan\frac{\sin[\omega_{0}t+\phi(t)]}{\cos[\omega_{0}t+\phi(t)]+\tan\frac{\theta}{2}\tan\theta_{+}(t)},
\end{split}
\end{equation}
and the contribution resulting from the geometric effect of the dynamical evolution, namely, the effective geometric phase, which can be expressed as
\begin{equation}
\label{eq:phaevo}
\begin{split}
   \Phi_{e}(t)&=i\int_{0}^{t}\Big\langle\Psi(\tau)\Big|\frac{\partial}{\partial\tau}\Big|\Psi(\tau)\Big\rangle d\tau\\
              &=\int_{0}^{t}[\omega_{0}-s(\tau)]\cos^{2}\theta_{+}(\tau)d\tau.
\end{split}
\end{equation}
The reason why the phase in Eq.~\eqref{eq:phaevo} is called the effective geometric phase is that $\Phi_{e}(t)$ is closely associated with the dynamical evolution and that for a given evolution time, the Pancharatnam relative phase $\Phi_{P}(t)$ in Eq.~\eqref{eq:phaove} maybe disappear and makes no contribution to the dynamical evolution.
For the case that the system evolves along a quasicyclic path with the evolution time $t=2\pi/\omega_{0}$, the expression of the geometric phase in Eq.~\eqref{eq:geopha3} is consistent with that obtained in Refs.~\cite{PhysRevLett.105.240406,PhysRevA83.052121}.
Obviously, in contrast to that in an equilibrium environment, the renormalization of the intrinsic energy of the system induced by the environmental nonequilibrium feature, namely, the frequency shift $s(t)$ gives an additional contribution to the geometric phase.

It is worth mentioning the situation when the environment is in equilibrium $(a = 0)$ and the evolution of the system is quasicyclic with time $t=2\pi/\omega_{0}$. In this case, there is no frequency shift and the decoherence factor $F(t)$ is real with zero argument.
As a consequence, the Pancharatnam relative phase in  Eq.~\eqref{eq:phaove} is zero and the geometric phase in Eq.~\eqref{eq:geopha3} only arises from the effective geometric phase of the dynamical evolution which can be reduced to
\begin{equation}
\label{eq:geopha4}
   \Phi_{g}(t)=\Phi_{e}(t)=\omega_{0}\int_{0}^{t}\cos^{2}\theta_{+}(\tau)d\tau.
\end{equation}
This expression returns to the well-known results obtained in Refs.~\cite{PhysRevA73.052103,PhysRevA74.042311,PhysRevA81.022120}.

The effective geometric phase in Eq.~\eqref{eq:phaevo} can be expressed, by making the correction, as~\cite{PhysRevA74.042311}
\begin{equation}
\label{eq:geopha4}
   \Phi_{e}(t)=\Phi_{e}^{U}(t)+\delta\Phi_{e}(t),
\end{equation}
where $\Phi_{e}^{U}(t)=\omega_{0}t\cos^{2}(\theta/2)$ denotes the unitary effective geometric phase with no influence from the environment, and $\delta\Phi_{e}(t)$ is the correction to the effective geometric phase made between the cases under nonunitary and unitary dynamical evolution.
For the case under unitary dynamics and evolution time $t=2\pi/\omega_{0}$, the effective geometric phase can be reduced to $\Phi_{e}^{U}=\pi(1+\cos\theta)$.
Combining Eqs.~\eqref{eq:phaevo} with~\eqref{eq:geopha4}, it indicates that the frequency shift $s(t)$ induced by the environmental nonequilibrium feature also has a significant impact on the correction to the effective geometric phase.

\section{Results and discussion}
\label{sec:resu}

In this section, we show the results of the geometric effect of the dynamical evolution in Markovian and non-Markovian regions induced by the nonequilibrium environment.
We mainly focus on the influence of the environmental nonequilibrium feature on the geometric effect of the quantum dynamics. For simplicity, we set $\omega_{0}=0$ and use the case for the environment in equilibrium ($a=0$) as a reference.
In this case, the effective geometric phase can indirectly reflect the correction to the unitary geometric phase.

\begin{figure}[ht]
 \centering
    \includegraphics[width=3.45in]{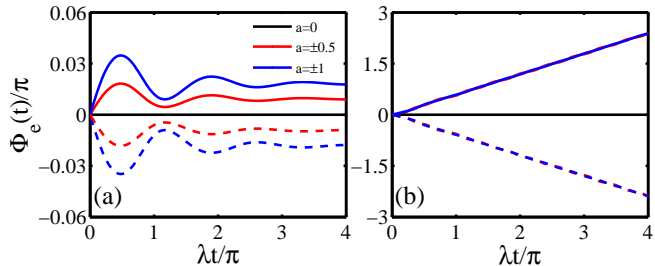}
    \caption{(Color online) Effective geometric phase $\Phi_{e}(t)$ as a function of evolution time $t$
    for different values of $a$ with $\theta=\pi/2$ in
    (a) Markovian dynamics region with $\nu=0.5\lambda$ and $\kappa=\lambda$
    and (b) non-Markovian dynamics region with $\nu=2\lambda$ and $\kappa=\lambda$
    (the solid and dashed lines are plotted for $a>0$ and $a<0$, respectively).}
    \label{Fig1}
\end{figure}

Figure~\ref{Fig1} shows the time evolution of the effective geometric phase $\Phi_{e}(t)$ for different environmental nonequilibrium parameter $a$ in Markovian and non-Markovian dynamics regions, respectively.
When the environment is out of equilibrium, in both dynamics regions, $\Phi_{e}(t)$ shows symmetrical behavior on opposite sides of the equilibrium case for positive and negative $a$.
As time goes by, in Markovian dynamics region as shown in Fig.~\ref{Fig1}(a), $\Phi_{e}(t)$ gradually tends to a stable value whereas it increases monotonically in non-Markovian dynamics region as shown in Fig.~\ref{Fig1}(b).
Furthermore, as the environment departs from equilibrium for a given evolution time, in Markovian dynamics region, $\Phi_{e}(t)$ deviates from the equilibrium case; on the contrary, the behavior of deviation is not obvious in non-Markovian dynamics region.
This suggests that the environmental nonequilibrium feature gives different additional contributions to the geometric effect of the dynamical evolution in the two dynamics regions and that the correction to the geometry of quantum evolution is mainly ruled by the nonequilibrium feature of the environment in Markovian dynamics region.
Furthermore, the effective geometric phase $\Phi_{e}(t)$ in non-Markovian dynamics region is much larger than that in Markovian dynamics region,
which indicates that the non-Markovian behavior in the dynamics makes large correction to the geometric effect of the dynamical evolution. This shows very good agreement with the conclusions obtained in Ref.~\cite{PhysRevA91.042111}.

\begin{figure}[ht]
 \centering
    \includegraphics[width=3.45in]{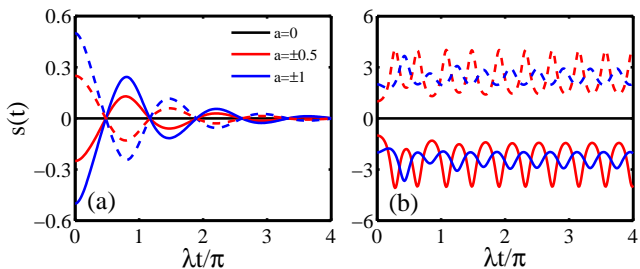}
    \caption{(Color online) Time evolution of the frequency shift $s(t)$ for different values of $a$
    with $\theta=\pi/2$ and $\kappa=\lambda$ in
    (a) Markovian dynamics region with $\nu=0.5\lambda$
    and (b) non-Markovian dynamics region with $\nu=2\lambda$
    (the solid and dashed lines are also for positive and negative $a$, respectively).}
    \label{Fig2}
\end{figure}

To study the reason of difference in the effective geometric phase between the equilibrium and nonequilibrium cases, we show the frequency shift $s(t)$ for different values of $a$ in Markovian and non-Markovian dynamics regions in Fig.~\ref{Fig2}(a) and (b), respectively.
In both dynamics regions, $s(t)$ shows the symmetry for $a$ taking positive and negative values.
When the environment is in nonequilibrium, in Markovian dynamics region, $s(t)$ decays with oscillatory behavior and discrete zeros and it asymptotically approaches zero as time goes on. However, $s(t)$ oscillates periodically in time with nonzero midline in non-Markovian dynamics region.
Furthermore, as the environment deviates from equilibrium for a given evolution time, $s(t)$ increases in Markovian dynamics region whereas it hardly changes in non-Markovian dynamics region.
The behavior in time-dependent frequency shift $s(t)$ is closely associated with the effective geometric phase as shown in Fig.~\ref{Fig1}. It further indicates that the environmental nonequilibrium feature which gives rise to the renormalization of the intrinsic energy of the quantum system plays an important role in the geometric effect of the dynamical evolution.

\begin{figure}[ht]
 \centering
    \includegraphics[width=3.45in]{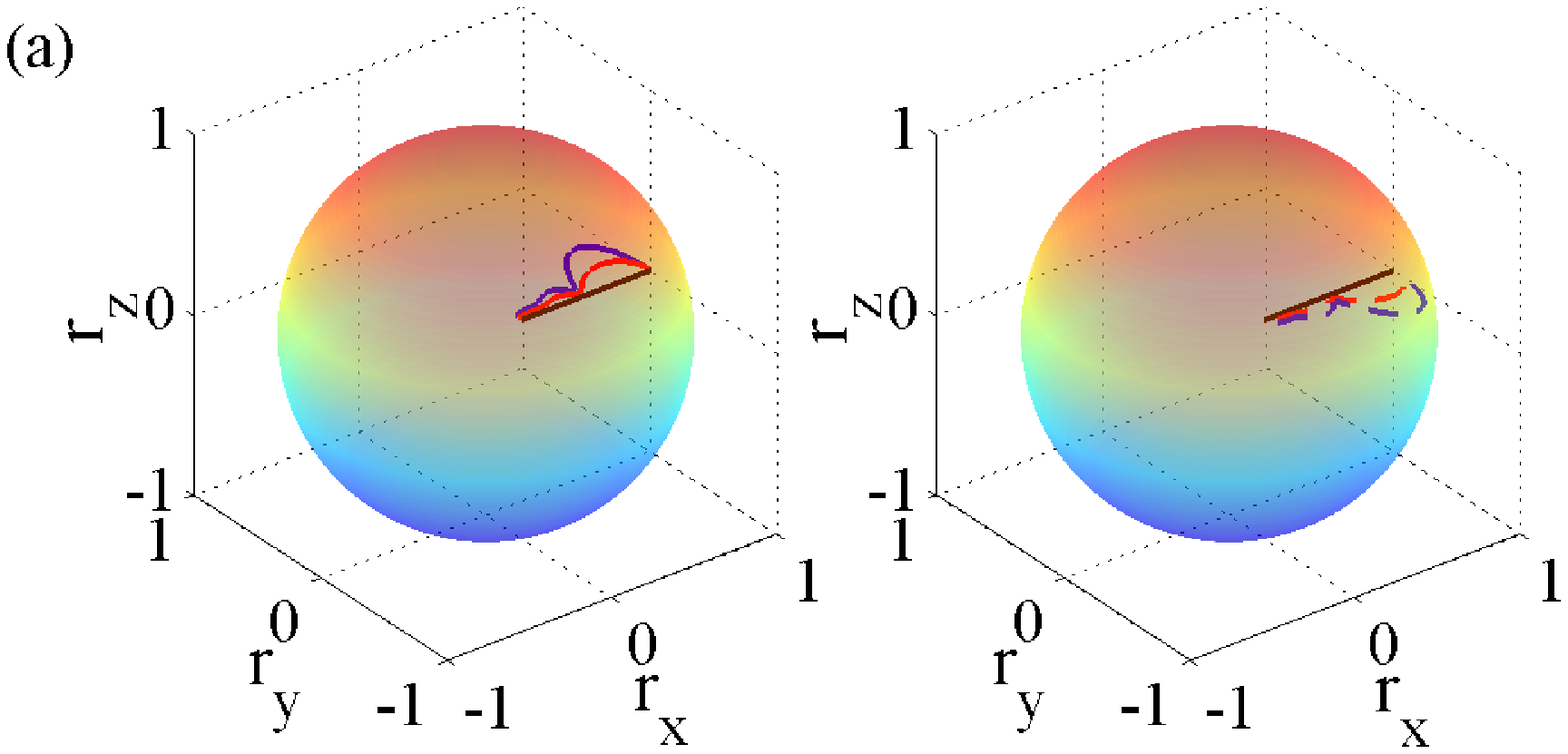}
    \includegraphics[width=3.45in]{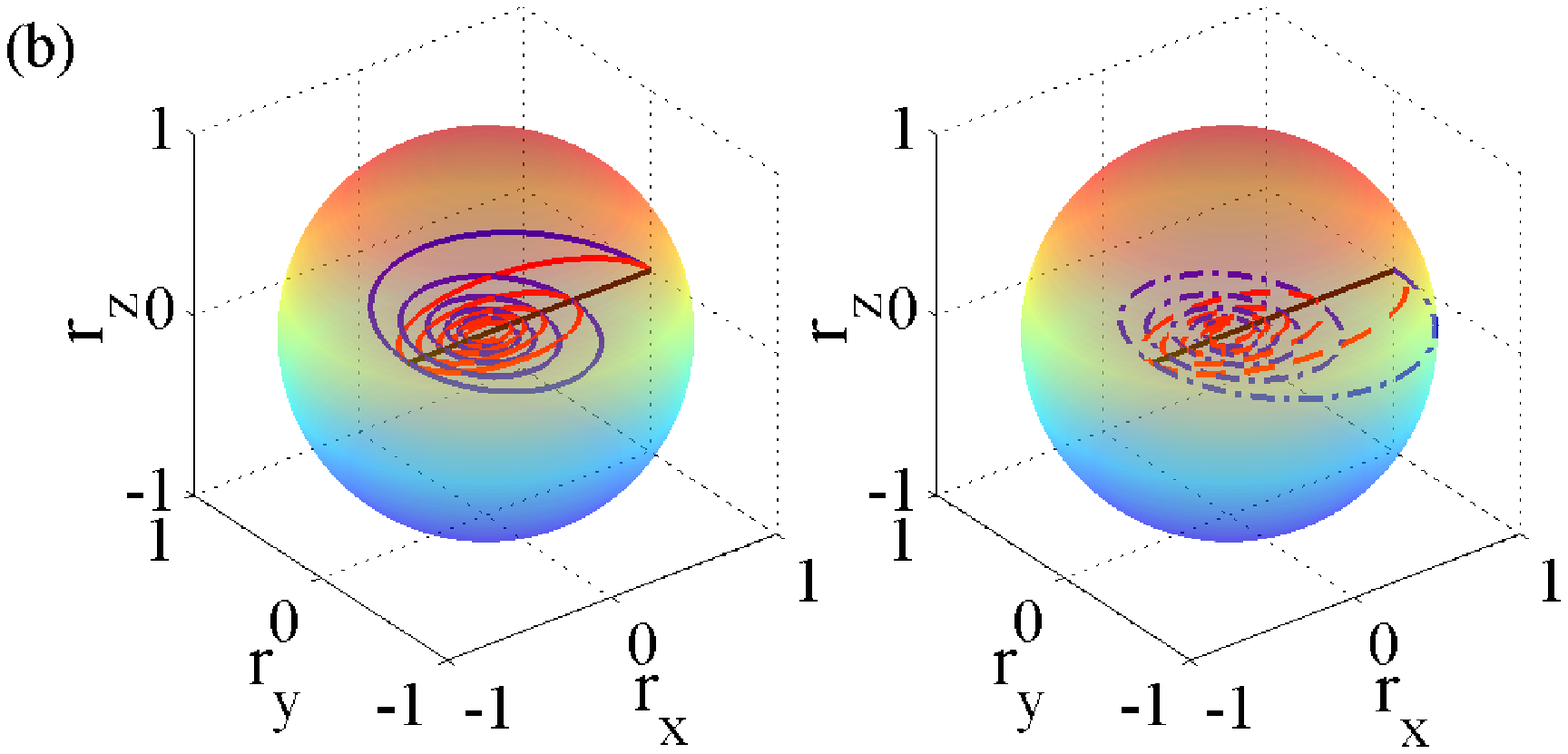}
    \caption{(Color online) Geometry of the dynamical evolution of the quantum system in Bloch sphere representation：
    (a) Markovian dynamics region with $\nu=0.5\lambda$ and $\kappa=\lambda$
    and (b) non-Markovian dynamics region with $\nu=2\lambda$ and $\kappa=\lambda$.
    The left and right columns are plotted for $a>0$ and $a<0$, respectively.
    Blue line for $a=\pm1$, red line for $a=\pm0.5$ and black line for $a=0$.}
    \label{Fig3}
\end{figure}

In order to study how the environmental nonequilibrium feature influences the path of the dynamical evolution, we show the time evolution of the reduced density matrix of the system for different values of $a$ in the Bloch sphere representation in Fig.~\ref{Fig3}.
This helps us to understand better the combined effects of the unitary and nonunitary parts of dynamical evolution which are closely associated with the energy renormalization and dynamical decoherence.
The radius $|\vec{r}(t)|$ of the Bloch vector denotes the absolute value of the decoherence factor: the normal and tangential slopes of the radius are related to the decoherence rate and frequency shift, respectively, and the rate of change of the radius is associated with non-Markovian behavior in the system dynamics.
Obviously, when the environment is out of equilibrium, the geometric evolution path displays antisymmetrical behavior in the Bloch sphere for positive and negative $a$ in both dynamics regions.
In Markovian dynamics region as shown in Fig.~\ref{Fig3}(a), $|\vec{r}(t)|$ decays monotonically whereas as shown in Fig.~\ref{Fig3}(b) in non-Markovian dynamics region, it decays with periodical oscillations, namely coherence revivals induced by environmental backaction.
Furthermore, in non-Markovian dynamics region, the oscillatory behavior in $|\vec{r}(t)|$ gets reduced as the environment departs from equilibrium, which reflects that the environmental nonequilibrium feature can suppress non-Markovian behavior in the system dynamics.
Moreover, as the environment deviates from equilibrium in both dynamics regions, the length of evolution path becomes longer which suggests that the environmental nonequilibrium feature reduces the dynamical decoherence of the quantum system.

\section{Conclusions}
\label{sec:conl}

In this paper, we have studied the geometry of dynamical evolution of a two-level quantum system coupled to a nonequilibrium noisy environment.
Due to the nonstationary statistical properties of the environmental noise, the decoherence factor is a complex time-dependent function and the imaginary part of the decoherence factor gives an additional contribution to the unitary evolution of the system dynamics.
Based on the quantum master equation in a nonequilibrium environment, we derived the time evolution of the geometric phase closely associated with the renormalization of the intrinsic energy of the system, namely, the frequency shift.
We have demonstrated that the environmental nonequilibrium feature plays a crucial role in both the geometric phase and evolution path of the quantum dynamics.
It was shown that the nonequilibrium feature of the environment makes the length of evolution path becomes longer and reduces the dynamical decoherence of the quantum system compared with the equilibrium case.
This result is significant to quantum information processing based on the geometry of dynamical evolution of open quantum systems.

The investigation on the geometric effect of dynamical evolution in a nonequilibrium environment helps us understand better the non-Markovian decoherence dynamics of open quantum systems.
Within some theoretical and experimental frameworks, the phase information of quantum evolution
can be measured by the interferometric measurement via nuclear magnetic resonance (NMR)
or by the current measurement via a QPC device~\cite{PhysRevLett.105.240406,PhysRevLett.91.100403,PhysRevA76.042121,PhysRevB96.235417,SciRep.5.11726}.
In principle, the observation of the environmental nonequilibrium feature on the geometry of dynamical evolution would be expected to be realized experimentally by using a NMR interferometry
or a QPC detector based on the theoretical frameworks demonstrated in Refs.~\cite{PhysRevLett.105.240406} and~\cite{SciRep.5.11726}, respectively.

\begin{acknowledgments}
This work was supported by the Natural Science Foundation of Shandong Province under Grant Nos. ZR2014AM030 and ZR2016AP14.
X.C. and R.M. also acknowledge the support from the Doctoral Research Fund of Shandong Jianzhu University (Grant Nos. XNBS1852 and XNBS1860).
\end{acknowledgments}


%

\end{document}